\documentstyle[twocolumn,psfig,floats,aps,amsmath,amssymb]{revtex}
\begin{document}
\twocolumn[\hsize\textwidth\columnwidth\hsize\csname
@twocolumnfalse\endcsname
\title{Thermodynamic Properties of the Spin-1/2 Antiferromagnetic
ladder Cu$_2$(C$_2$H$_{12}$N$_2$)$_2$Cl$_4$ under Magnetic Field
}
\author{R.~Calemczuk$^a$, J.~Riera$^{b,c}$, D.~Poilblanc$^b$,
J.-P.~Boucher$^{d}$, G.~Chaboussant$^e$, L.~Levy$^{e}$ and O.~Piovesana$^f$
}
\address{
$^a$DRFMC, 
Service de Physique Statistique, Magn\'etisme et Supraconductivit\'e,
CENG, F-38054 Grenoble cedex 9, France\\
$^b$Laboratoire de Physique Quantique \& UMR--CNRS 5626,
Universit\'e Paul Sabatier, F-31062 Toulouse, France\\
$^c$Instituto de F\'{\i}sica Rosario, Consejo Nacional de 
Investigaciones 
Cient\'{\i}ficas y T\'ecnicas y Departamento de F\'{\i}sica\\
Universidad Nacional de Rosario, Avenida Pellegrini 250, 2000-Rosario,
Argentina\\
$^d$Laboratoire de Spectrom\'etrie Physique, Universit\'e Joseph Fourier,
BP 87, F-38402 St Martin d'H\`eres, France\\
$^e$ Grenoble High Magnetic Field Laboratory, CNRS and MPI-FKF, BP 166,
F-38042 Grenoble, France. \\
$^f$Dipartimento di Chimica, Universit\`a di Perugia, I-06100 Perugia, Italy
}
\date{\today}
\maketitle
\begin{abstract}

Specific heat ($C_V$) measurements in the spin-1/2 
Cu$_2$(C$_2$H$_{12}$N$_2$)$_2$Cl$_4$ system under a magnetic field
up to $H=8.25 T$ are reported and compared 
to the results of numerical calculations based on the 2-leg
antiferromagnetic Heisenberg ladder. While 
the temperature dependences of both the susceptibility and the low field
specific heat are accurately reproduced by this model,
deviations are observed below the critical field $H_{C1}$ at which 
the spin gap closes.
In this Quantum High Field phase, the contribution of the
low-energy quantum fluctuations are stronger than in the Heisenberg 
ladder model.
We argue that this enhancement can be attributed to dynamical lattice 
fluctuations. Finally, we show that such a  Heisenberg 
ladder, for $H>H_{C1}$, is unstable, when coupled to the 3D lattice, against a
lattice distortion. 
These results provide an alternative explanation for the observed
low temperature ($T_C\sim 0.5K$ -- $0.8K$) phase
(previously interpreted as a 3D magnetic ordering) as 
a new type of incommensurate gapped state.
\smallskip


\end{abstract}

\vskip2pc]
Wide interest is currently devoted to "gapped" spin systems,
both experimentally and theoretically. In one dimension, $S=1$ Haldane
and alternating spin chains provide good examples of such systems.
When magnetic frustration is present, similar situations can be
found not only in one-dimension (1D) -- the $J_1-J_2$ model for
instance -- but also in two (2D) or three dimensions. An intermediate
situation between 1D and 2D is provided by the so-called
"ladder" systems, which couple an even number of quantum spin
($S=1/2$) chains. As for alternating and frustrated spin chains,
the energy diagram is characterized by an energy gap
$\Delta_S$ between the $S=0$ ground state (GS) and the first magnetic
$S=1$ excited state
leading to characteristic magnetic and thermodynamic
properties at low enough temperature, $T<\Delta_S$.

The application of a magnetic field yields drastic changes in the
energy spectrum. In particular, as a result of the Zeeman splitting
undergone by the $S=1$ excited state, a second-order transition
occurs at the critical field $H_{C1}=\Delta_S/g\mu_B$, where $g$ is
the gyromagnetic ratio and $\mu_B$ the Bohr magneton. Above
$H_{C1}$, the GS becomes magnetic~\cite{IA} and a continuum
of excitations develops giving rise to "incommensurate" zero-energy
fluctuations.

\begin{figure}[htb]
\begin{center}
\vspace{-0.8truecm}
\psfig{figure=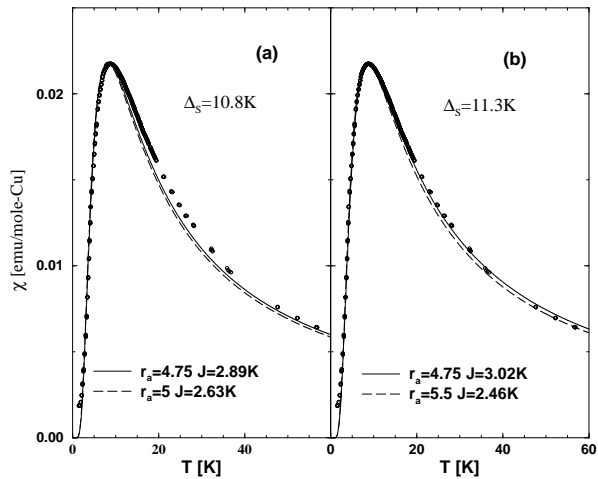,width=8truecm,angle=-90}
\end{center}
\caption{
Theoretical fit of the temperature dependence of the magnetic
susceptibility. The experimental data are taken from 
Refs.~\protect\cite{Chaboussant0,Chaboussant1}. 
The anisotropy ratios $r_a=J_\perp/J_\parallel$ as well as the
magnetic coupling along the chain $J=J_\parallel$ are indicated on the plots.
(a) and (b) correspond to various set of parameters producing
two different spin gaps $\Delta_S$. 
}
\label{suscep}
\end{figure}

Experimentally, the study of such
a Quantum High Field (QHF) phase, i.e. for $H>H_{C1}$,
requires to work on systems having a relatively small gap $\Delta_S$.
Indeed, insulating ladders such as SrCu$_2$O$_3$~\cite{SrCuO} 
whose structure is closely related to the parent 2-dimensional cuprate 
antiferromagnets typically have spin gaps larger than $100K$. 
This explains that such a phase has rarely been investigated.
As shown by recent studies, an interesting opportunity is provided
by the compound Cu$_2$(C$_2$H$_{12}$N$_2$)$_2$Cl$_4$
(also known as CuHpCl) which is thought to behave as an ideal
2-leg spin-1/2 ladder system with a critical field of
$H_{C1}\simeq 7.5T$. Magnetic measurements have been used first to
characterize the magnetic parameters of the spin system.
The behavior of the magnetic susceptibility, reproduced from
Refs.~\cite{Chaboussant0,Chaboussant1} in Fig.~\ref{suscep},
is consistent with a gap of the order 
of $\Delta_S\sim 11K$.
Specific heat ($C_V$) measurements 
in CuHpCl have recently been performed under
a magnetic field of up to $9T$~\cite{Hammar}. 
In low field ($H<H_{C1}$), a single maximum is observed 
at relatively high temperature ($T\sim J_\perp$), and, due to the 
presence of the energy gap, $C_V$ decreases exponentially at low temperature 
[$\sim \exp{(-\Delta_S/T)}$]. 
In addition, a second order transition was shown to
occur at very low temperature ($0.5K <T_C< 0.8K$)
and was interpreted as the onset of 3-dimensional (3D) magnetic order.

In the present work,
new specific heat measurements in a field (up to $8.25T$) are presented which
mainly focus on the QHF phase. 
They were performed
by using a scanning adiabatic method. A small known power is applied to a
high purity silicon sample holder and the temperature difference between the
sample and a surrounding radiation screen is measured by a gold-iron
thermocouple using a DC squid device as a current amplifier. A feedback
network maintains the radiation screen at the same temperature as the
sample, then strongly reducing the heat exchange process. The temperature
rises slowly from 0.1 K to 8K, at a speed as slow as 10 mK/min. The
measurement of the temperature of the radiation screen then allows the
specific heat of the sample to be calculated. Such a slow drift rate in
temperature ensures that all parts of the sample are in thermal equilibrium,
and unlike pulsed methods the specific heat of non metallic materials with
poor thermal diffusivities can be accurately measured. In the present work,
four single crystals glued onto a mica plate (as in previous susceptibility
experiments [3,4]) were measured. Each crystal weighed approximately 0.5 mg.
The contributions of all the addenda - the mica , varnish, vacuum grease,
and the silicon sample holder - were estimated and subtracted. 
The behaviors observed are
directly compared to the results of a numerical investigation
based on the Heisenberg ladder model. 
In the QHF phase, a second maximum develops at low temperature.
Above $T_C$, we observed deviations from the isolated ladder model 
(the contribution of the low energy "incommensurate" spin fluctuations 
occurs at larger fields)
which, we argue, can be due to dynamical lattice fluctuations.
We suggest an alternative explanation for the low temperature 
phase in terms of a new incommensurate gapped state. 
A calculation based on a Heisenberg ladder coupled to 3D (classical)
phonons strongly supports this scenario. 

The Hamiltonian we shall use to describe the compound is the Heisenberg 
model on a ladder defined by 
\begin{eqnarray}
   {\cal H}=
   J_\perp \; \sum_{j} 
   {\bf S}_{j,1} \cdot {\bf S}_{j,2} 
   +J_\parallel\;\sum_{\beta,j} 
   {\bf S}_{j,\beta} \cdot {\bf S}_{j+1,\beta}
\label{hamiltonian} 
\end{eqnarray}
\noindent
where  $\beta$ (=1,2)
labels the two legs of the ladder (oriented along the 
$x$-axis), $j$ is a rung index ($j$=1,...,$L$) and $J_\parallel$ and 
$J_\perp$ are
the bond strengths along and between the chains respectively. 
An applied field $H$ in the $Z$-direction leads to an additional
Zeeman term, ${\cal H}_Z=-g \mu_B H \sum_{\beta,j} S^Z_{j,\beta}$,
with an average value $g\simeq 2.08$~\cite{Chaboussant1}.

Our numerical approach is based on Exact Diagonalisation 
techniques. At $T=0$,
clusters with size up to $2\times 14$ can be handled with the Lanczos 
algorithm allowing, after a proper finite size scaling procedure, 
for accurate determinations of the various physical 
quantities~\cite{note_gap}. At finite temperature, 
a full diagonalization of $2\times 6$, $2\times 8$ and
$2\times 10$ ladders has been performed. According to previous literature,
the anisotropy ratio $r_a=J_\perp/J_\parallel$ lies around $5$.
In this regime, the spin correlation length is
smaller than the system sizes so that finite 
size corrections become negligable.
 
\begin{figure}[htbp]
\begin{center}
\vspace{-1.0truecm}
\psfig{figure=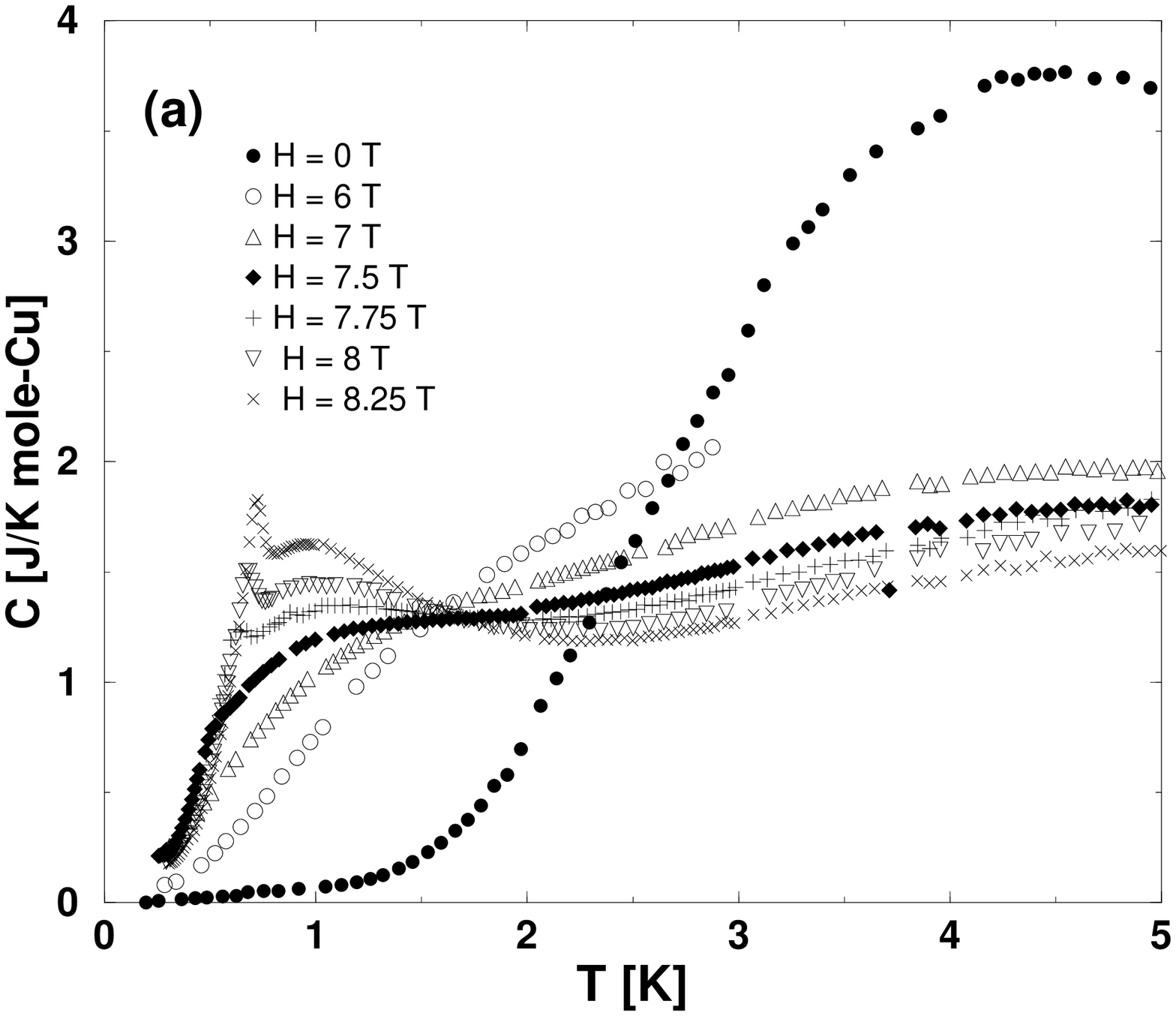,width=8truecm,angle=-90}
\vspace{-0.75truecm}
\psfig{figure=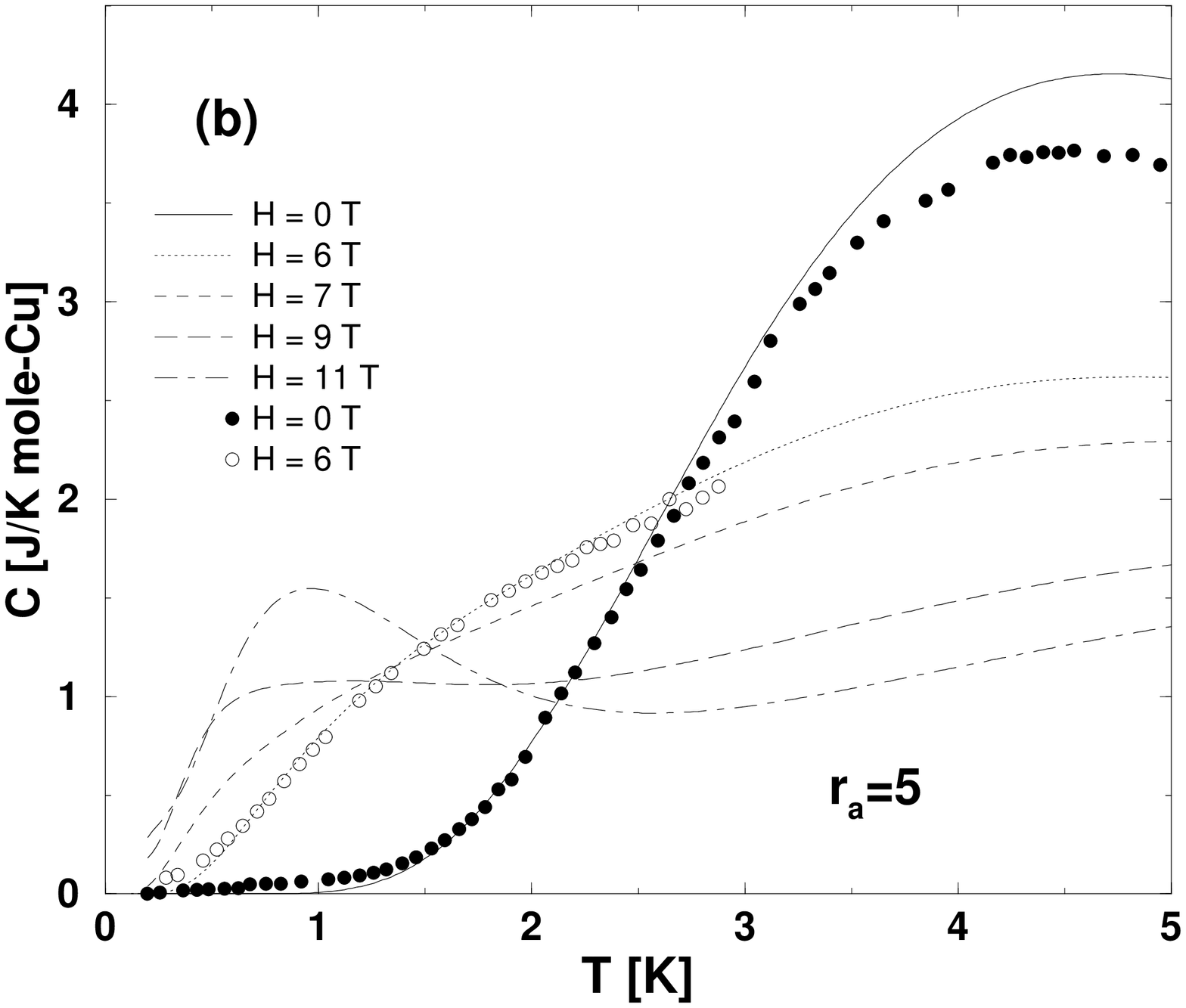,width=8truecm,angle=-90}
\vspace{-0.75truecm}
\psfig{figure=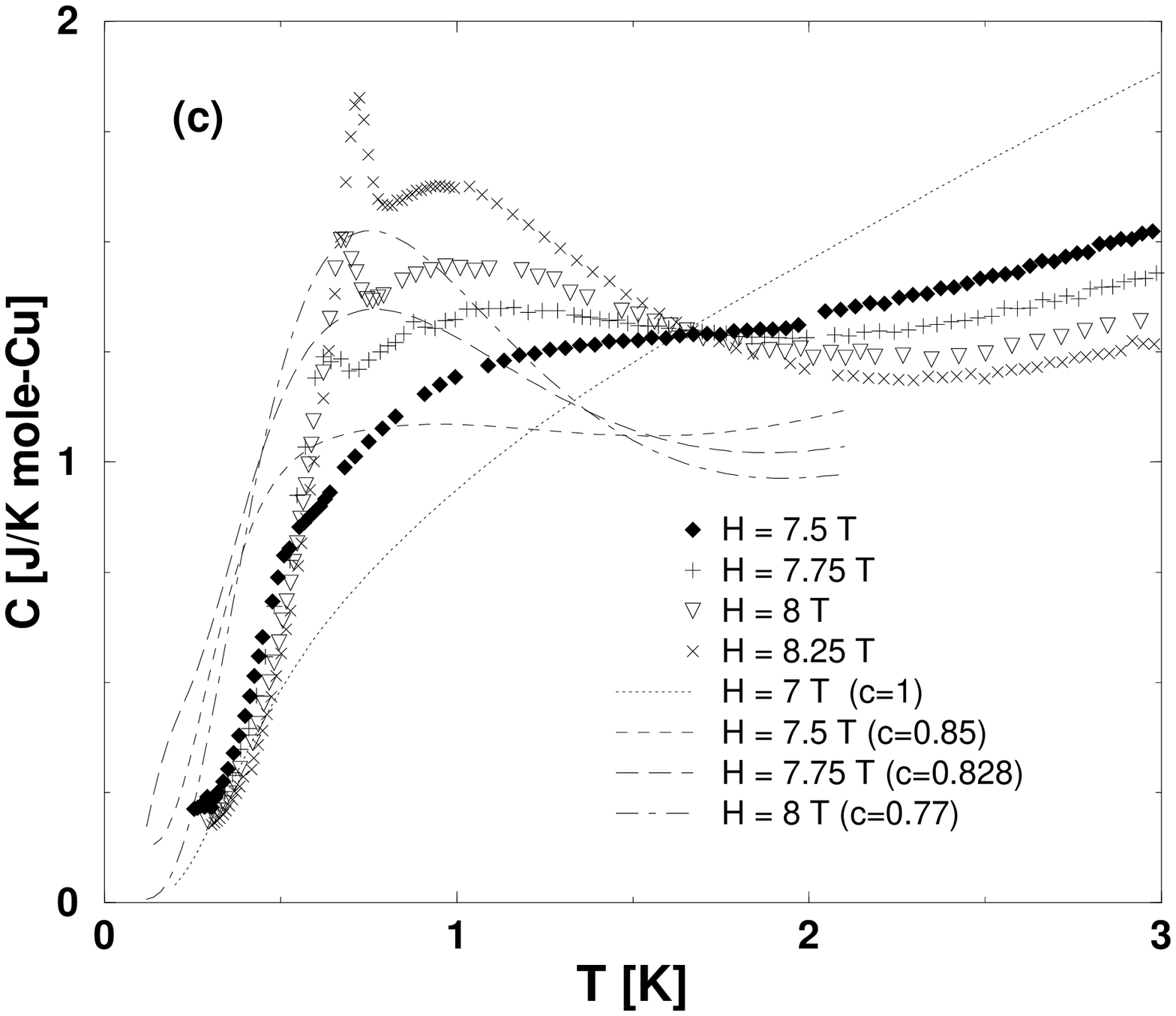,width=8truecm,angle=-90}
\end{center}
\caption{
Specific heat vs T for different values of the magnetic 
field.
(a) Experimental data.
(b,c) Comparison between experiment (symbols) and theory (lines)
using $r_a=5$, $J_\parallel=2.63K$.
}
\label{cv}
\end{figure}

In order to test the choice of parameters in (\ref{hamiltonian}), let us
briefly consider the temperature dependence of the magnetic
susceptibility~\cite{Chaboussant1}.
A comparison between the numerical and the experimental data 
is shown in Fig.~\ref{suscep}. 
In fact, the quality of the fit is not very sensitive to the anisotropy 
ratio (in the range $4.75\le r_a\le 5.5$).
Parameters producing a gap of $\Delta_S\simeq 10.8K$ (as reported in
Ref.~\cite{Chaboussant1}) or $\Delta_S\simeq 11.3$ (as reported in
Ref.~\cite{Chaboussant2}) give excellent fits of the experimental data.
As shown numerically~\cite{Hayward}
or in a strong coupling approximation~\cite{Chaboussant2}
the experimental behavior of the magnetization vs H is well reproduced
by a similar set of parameters.

Concerning the specific heat measurements shown in Figs.~\ref{cv},
the exponential behavior at low temperature characteristic of
the spin gap is suppressed at moderate magnetic fields.
Above $7.5T$, a broad maximum in $C_V(T)$
builds up signaling the emergence of new low energy 
fluctuations\cite{Mohan}. 
The numerical calculations of $C_V(T)$ 
in Fig.~\ref{cv}(b) based on the above ladder
model (with parameters leading to $H_{C1}\simeq 7.7 T$) 
reveal qualitatively the same behavior. 
At low field, up to $H=6T$, the agreement with experiment is very good,
hence establishing the relevance of the ladder model (\ref{hamiltonian})
in this regime.
In the QHF, however, the maximum observed in the theoretical calculation
appears at higher magnetic fields than in experiment. Deviations 
from the theoretical behavior appear at low temperature 
for fields above $7.5T$ after the closing of the ladder gap. 
We argue here that this effect can be due to lattice fluctuations.
In fact, it has been shown~\cite{Sandvik} for uniform Heisenberg chains,
in the context of spin-Peierls transitions, 
that an underlying spin-lattice coupling
can lead to significant deviations e.g. in the magnetic susceptibility
which can be accounted for by an effective exchange coupling.
As shown in Fig.~\ref{cv}(c), 
a behavior qualitatively similar to the experimental observations 
can be obtained by using renormalized exchange couplings
$J^{\rm eff}_\mu=c\, J_\mu$ 
($\mu=\perp,\parallel$), $c\le 1$, which, according to Ref.~\cite{Sandvik}, 
is consistent with the effect of a coupling to the lattice.
For increasing field above $7T$, the renormalization parameter $c$
decreases signaling an increasing role played by the lattice coupling.

Motivated by the above discussion, we reasonably assume the presence
of a magneto-elastic 
coupling along the chain direction~\cite{note3} by replacing 
the second term of Eq.~(\ref{hamiltonian}) by 
\begin{eqnarray}
   {\cal H}_\parallel=
   J_\parallel\;\sum_{\beta,j} (1+\delta_{\beta,j})
   {\bf S}_{j,\beta} \cdot {\bf S}_{j+1,\beta} 
+ \frac{1}{2}K \sum_{\beta,j} \delta_{\beta,j}^2 \, ,
\label{ham2} 
\end{eqnarray}
\noindent
where the second term corresponds to the (3D) lattice elastic energy 
and where the set of parameters $\{\delta_{\beta,j}\}$ (proportional to the 
atomic displacements) have to be determined by minimizing the total energy.

Hamiltonian (\ref{ham2}) can lead to a lattice distortion 
(i.e. that $\{\delta_{\beta,j}\}\ne 0$)
in the strong coupling limit i.e. 
when $J_\perp\gg J_\parallel$. In this case, for $H\ge H_{C1}$,
by retaining the $S^Z=1$ and $S^Z=0$ states only on each rung 
(see e.g.~\cite{Chaboussant2,Mila}), the spin
ladder reduces to a 1D spinless fermion model~\cite{Haldane} 
with a hopping amplitude $t=J_\parallel/2$
and a nearest neighbor repulsion $V=J_\parallel/2$.
Physically, a particle corresponds, in the original spin language, 
to a $S_Z=1$ rung triplet excitation so that the effective band filling is 
directly proportional to the relative magnetization $M/M_{\rm sat}$ 
in such a way that $2k_F=2\pi (M/M_{\rm sat})$. 
If one neglects the short range repulsion $V$ between the particles
(this should be justified for low particle 
density i.e. small magnetization), as in the usual Peierls transition,
a modulation of the hopping amplitude $t=J_\parallel/2$ of wavevector
$2k_F$ and magnitude $\delta$ opens a gap at the chemical 
potential and leads to
an energy gain $\Delta E\propto \delta^2 \ln{({\rm const}/\delta)}$, 
for $\delta\ll 1$. For arbitrary large $K$, the minimum of the total 
energy is then 
obtained for an equilibrium value $\delta\sim \exp{(-{\rm const}K)}$.

\begin{figure}[htb]
\begin{center}
\vspace{-0.8truecm}
\psfig{figure=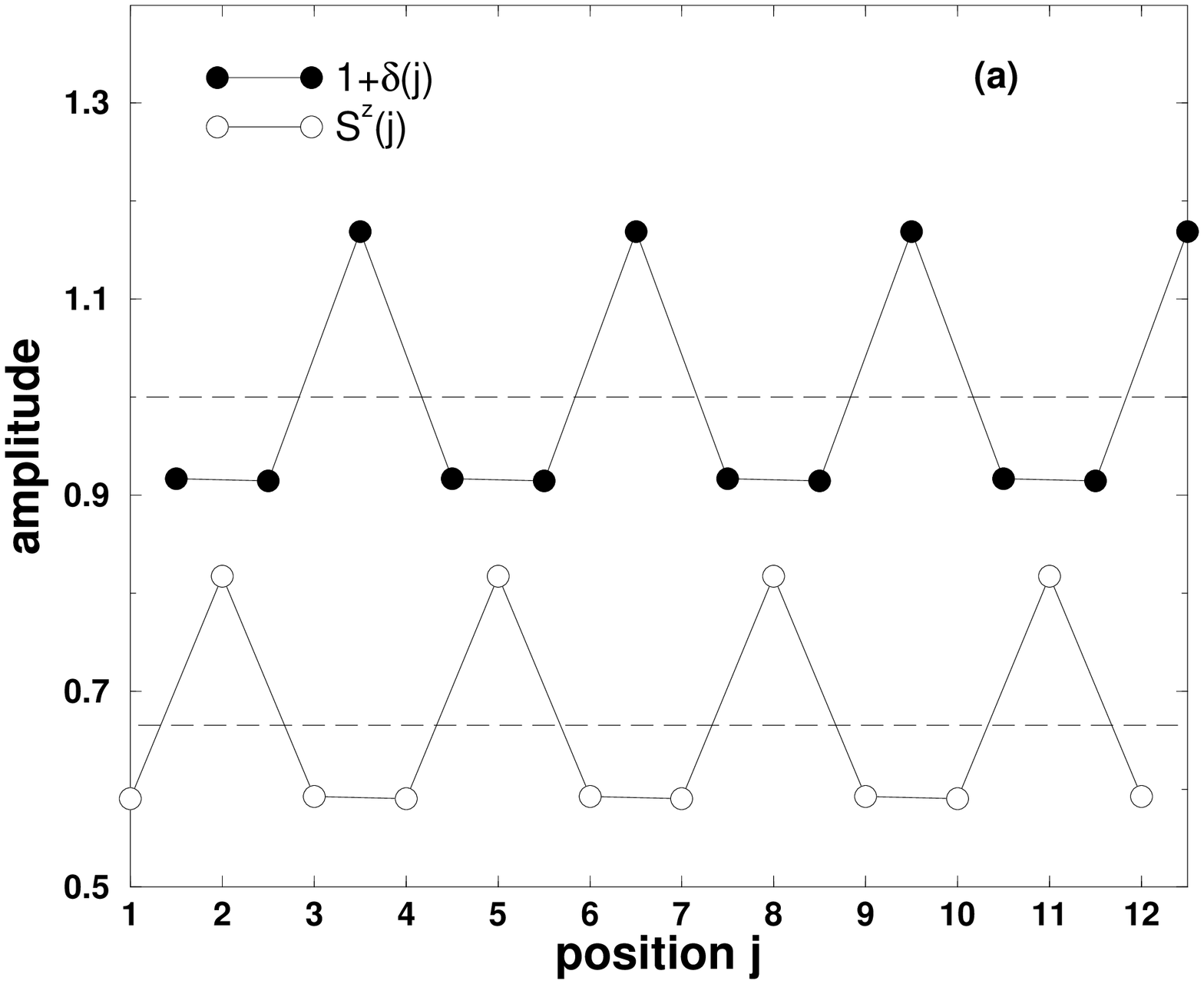,width=8truecm,angle=-90}
\vspace{-0.65truecm}
\psfig{figure=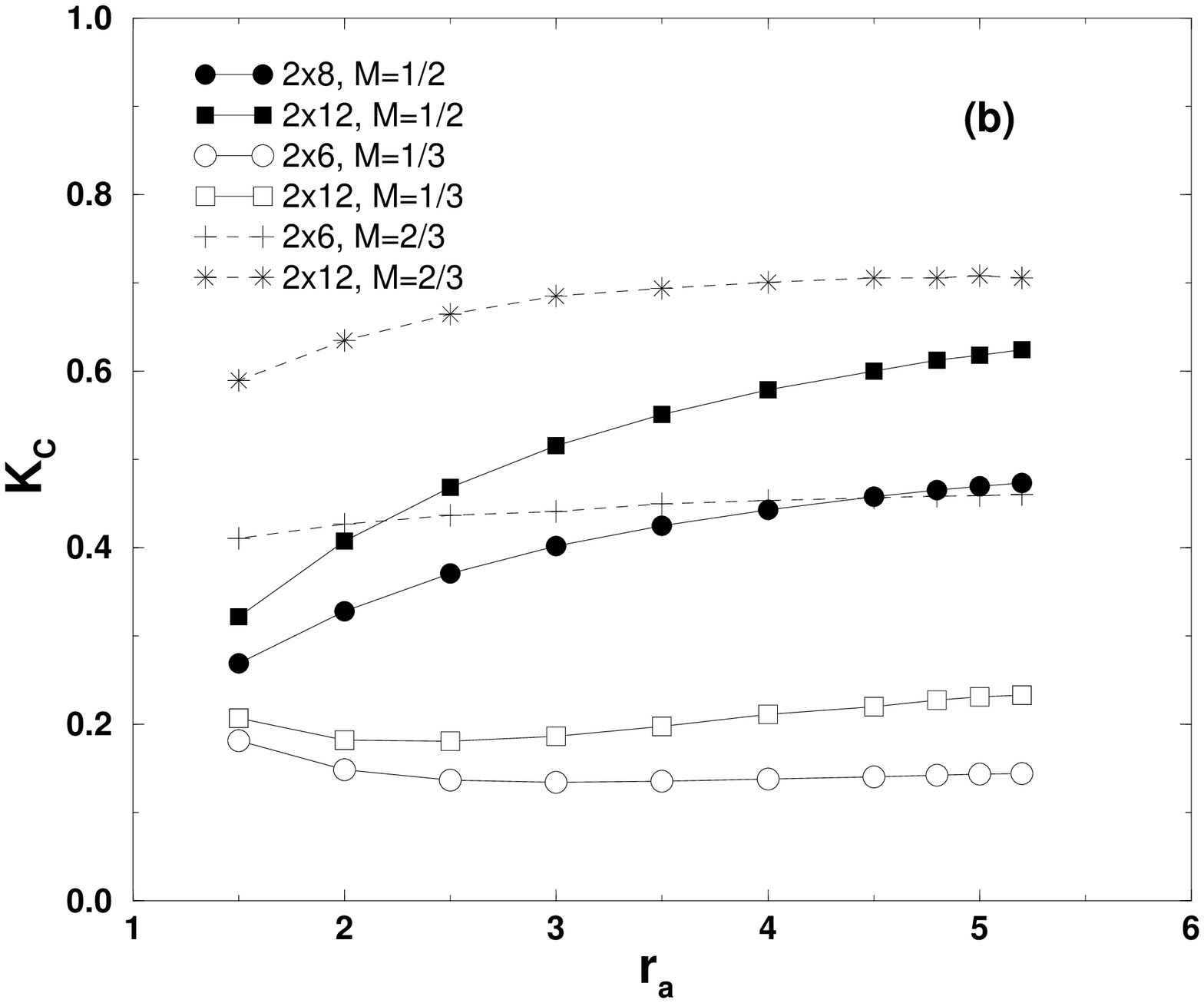,width=8truecm,angle=-90}
\end{center}
\caption{
(a) Equilibrium modulation $1+\delta_j$ and corresponding
average spin density $\big< (S_{j,1}^Z+S_{j,2}^Z) \big>$ vs
the position along the ladder direction calculated on a $2\times 12$
cluster for $r_a=5$, $K=0.6$ and $M/M_{\rm sat}=2/3$.
(b) Critical value $K_C$ as a function of the anisotropy ratio
$r_a=J_\perp/J_\parallel$ for $M/M_{\rm sat}=1/3$,
$1/2$ and $2/3$. The system sizes are indicated on the plot.
}
\label{peierls}
\end{figure}

In order to explore the relevance of the previous scenario,
a numerical investigation 
on finite clusters is required. The following study has been 
restricted to $T=0$ and simple ratios for
$M/M_{\rm sat}$ like 1/3, 1/2 or 2/3. The minimization of the total
energy using expression (\ref{ham2}) can be realized by an 
Exact Diagonalization technique supplemented by a 
self-consistent procedure~\cite{Riera}.
A typical GS configuration is shown in Fig.~\ref{peierls}(a) for a 
magnetization $M=\frac{2}{3}M_{\rm sat}$. 
The lowest energy is obtained for a perfectly symmetric
modulation of the two chains i.e. $\delta_{\beta,j}=\delta_j$.
The variation of the 
distortion $\delta_j$ along the ladder is correlated with that of
the spin density and is a periodic function of period 
$\lambda_F=M_{\rm sat}/M$. 
For the special case $M/M_{\rm sat}=1/2$, the distortion becomes
commensurate and corresponds to 
a simple dimerization of the lattice, 
similar to the D-phase observed in spin-Peierls 
chains such as CuGeO$_3$ in the {\it absence} of a magnetic field. 

In order to study the stability of these modulated phases
we have defined a critical elastic constant as
$K_C=\lim_{|\delta_j|\rightarrow 0} \{ \Delta E/\sum_j \delta_j^2\}$,  
where $\Delta E$ corresponds to the magnetic energy gain
due to the equilibrium distortion pattern.
The distorted phase is then stable for $K\le K_C$.
Our results displayed in Fig.~\ref{peierls}(b) as a function
of the ratio $J_\perp/J_\parallel$ show that $K_C$ increases with
system size; although it is difficult to extrapolate our
results to the thermodynamic limit, they clearly establish
that a small spin-lattice coupling leads to modulated structures.
It is interesting to notice that, for a small $M/M_{\rm sat}$ (i.e. for $H$
just above $H_{C1}$), a simple $2k_F$ modulation is expected
(in this regime, the short range repulsion $V$
becomes irrelevant) while,
for $M/M_{\rm sat}$ close to $1/2$ more complicated incommensurate
structures of ``soliton lattice'' type (i.e. involving an infinite number
of higher harmonics)~\cite{soliton} should be stabilized.
All these incommensurate structures bear strong similarities with the
I-phase of the spin-Peierls systems~\cite{Riera} and have similar properties
as e.g. the existence of a gap~\cite{gaps_IC}. 

To conclude, we have shown that the specific heat data 
for $H\le H_{C1}$ are quantitatively well described by an isolated Heisenberg
ladder model. Deviations from the predictions of this model observed
for $H> H_{C1}$ and low temperatures are attributed to 
the effect of a small magneto-elastic coupling to the 3D lattice.
We have shown numerically that, in this regime, the Heisenberg ladder
becomes unstable against a lattice distortion leading to
a new gapped incommensurate phase.

D.~P. and J.~R. thank IDRIS, Orsay (France) for
allocation of CPU time on the C94, C98 and T3E Cray supercomputers. 
J.~R. acknowledges partial support from the Ministry of Education (France) and 
the Centre National de la Recherche Scientifique.

\end{document}